\documentclass[12pt,preprint]{aastex}

\newcommand{\ms}{\mbox{m s$^{-1}$}}
\newcommand{\degrees}{\mbox{$^\mathrm{o}$}}

\begin{document}

\title{Hot Jupiter Variability in Eclipse Depth}

\author{Emily Rauscher,\altaffilmark{1} Kristen Menou,\altaffilmark{1}
James Y-K. Cho\altaffilmark{2}, Sara Seager\altaffilmark{3} \& Bradley
M. S. Hansen\altaffilmark{4}}

\altaffiltext{1}{Department of Astronomy, Columbia University, 550
West 120th Street, New York, NY 10027, USA}

\altaffiltext{2}{Astronomy Unit, School of Mathematical Sciences,
Queen Mary, University of London, Mile End Road, London E1 4NS, UK}

\altaffiltext{3}{Dept. of Earth, Atmospheric, and Planetary sciences,
and Dept. of Physics, Massachusetts Institute of Technology, 54-1626,
77 Massachusetts Ave., Cambridge, MA, 02139}

\altaffiltext{4}{Department of Physics and Astronomy and Institute for
Geophysics and Planetary Physics, University of California, 475
Portola Plaza, Box 951547, Los Angeles, CA 90095, USA}

\begin{abstract}
Physical conditions in the atmospheres of tidally locked,
slowly rotating hot Jupiters correspond to dynamical circulation
regimes with Rhines scales and Rossby deformation radii comparable to
the planetary radii. Consequently, the large spatial scales of moving
atmospheric structures could generate significant photospheric
variability. Here, we estimate the level of thermal infrared
variability expected in successive secondary eclipse depths, according
to hot Jupiter turbulent ``shallow-layer'' models. The variability, at
the few percent level or more in models with strong enough winds, is
within the reach of Spitzer measurements. Eclipse depth variability is
thus a valuable tool to constrain the circulation regime and global
wind speeds in hot Jupiter atmospheres.
\end{abstract}

\section{Introduction}

The regime of circulation in hot Jupiter atmospheres may be unlike any
of the familiar cases in the Solar System.  Hot Jupiters are gaseous
giant planets found in close, circular orbits around their parent
stars, with periods on the order of a few days
\citep[e.g.,][]{Butler06}.  General arguments suggest that these
planets should be tidally locked \citep{Guillot1996, Rasio1996, Lubow1997,
Ogilvie2004}: their permanent day-sides are then continuously subject
to intense stellar irradiation, while night-sides are only subject to
modest internal energy fluxes.  Given such an uneven energetic
forcing, atmospheric winds would tend to redistribute heat around the
planet. The nature and efficiency of this redistribution process is
important in determining a variety of hot Jupiter observational
properties \citep[e.g.,][]{Seager2005,
Burrows2005,Fortney2005,Barman2005,Iro,Burrows2006,Fortney2006}.

Two groups have attempted to address the global atmospheric
circulation problem in hot Jupiter atmospheres, using different
approaches \citep{ShowGui,Cho2003,Menou03,C&S}. On dynamical grounds,
it was argued in \citet[][see also Showman \& Guillot 2002]{Menou03},
and explicitly shown via turbulent shallow-layer simulations in
\citet{Cho2003,Cho2006}, that tidally locked hot Jupiters occupy a
regime of circulation that is qualitatively different from that of
Solar System giant planets. The few bands and prominent circumpolar
vortices emerging in these simulations can be understood in terms of
the large Rhines scale and Rossby deformation radius in these
atmospheres
\citep[e.g.,][]{ChoPolvani96a,ChoPolvani96b}. \citet{Menou03}
suggested that the large resulting spatial scales of moving
circulation structures could lead to detectable hot Jupiter
variability.

Soon, interesting observational constraints will be placed on these
circulation regime arguments.  Three hot Jupiters have been detected
through infrared secondary eclipses with the Spitzer Space Telescope:
HD189733b \citep{Deming2006}, HD209458b \citep{Deming2005}, and TrES-1
\citep{Charbonneau2005}.  The planetary day-side flux is deduced from
the eclipse depth measured when the planet is hidden behind its
star. Repeated eclipse measurements could thus reveal detectable
levels of variability of the planetary day-side flux in these three
systems. In this Letter, we quantify the level of thermal infrared
variability expected in secondary eclipse depths, according to the
shallow-layer models of \citet{Cho2003,Cho2006}.

\section{Shallow-Layer Circulation Models}

As explained in detail in \citet{Cho2003,Cho2006}, turbulent
equivalent-barotropic models published to date greatly emphasize
dynamical aspects of hot Jupiter atmospheric circulation. We adopt the
same notation and adiabatic models as in \citet{Cho2006}: the global
wind strength, $\bar U$, and the net amplitude of radiative forcing,
$\eta$, are parameterized, not predicted. However, given these two
(and other relevant global planetary) parameters, the turbulent
atmospheric circulation is consistently found to develop a broad
equatorial wind and two large circumpolar vortices revolving around
the poles in several planetary days (= orbits). The dynamically
modified layer thickness in these models is a proxy for the planet's
photospheric temperature field. We refer the reader to \citet{Cho2006}
for details on the models, as well as a vast exploration of their
parameter space.

Our limited goal here is to show that the thermal variability expected
in at least some of these models is sufficiently large to be
detectable via repeated Spitzer secondary eclipse
measurements. Consequently, we focus on a limited set of four models,
with global wind speed values from $\bar U=100$ to $800$~\ms~and a
moderate amplitude of radiative forcing, $\eta =0.05$ (allowing the
weak thermal contrast of features in slow wind models to remain
apparent).

The models are explicitly calculated for HD209458b
parameters, at moderate T63 ($192 \times 96$ grid) resolution, over a
hundred planetary days or more. Resolution tests (up to T170) show
that T63 is sufficient to capture atmospheric temperature features
well enough for our present purpose. Daily outputs from the
simulations are used to generate model temperature maps of the
day-side thermal emission in our variability study (see
below). Table~1 summarizes the range of photospheric temperatures
derived from these four models, for the two brightest planets in our
study.

\begin{figure}
\begin{center}
\includegraphics[width=0.35\textwidth]{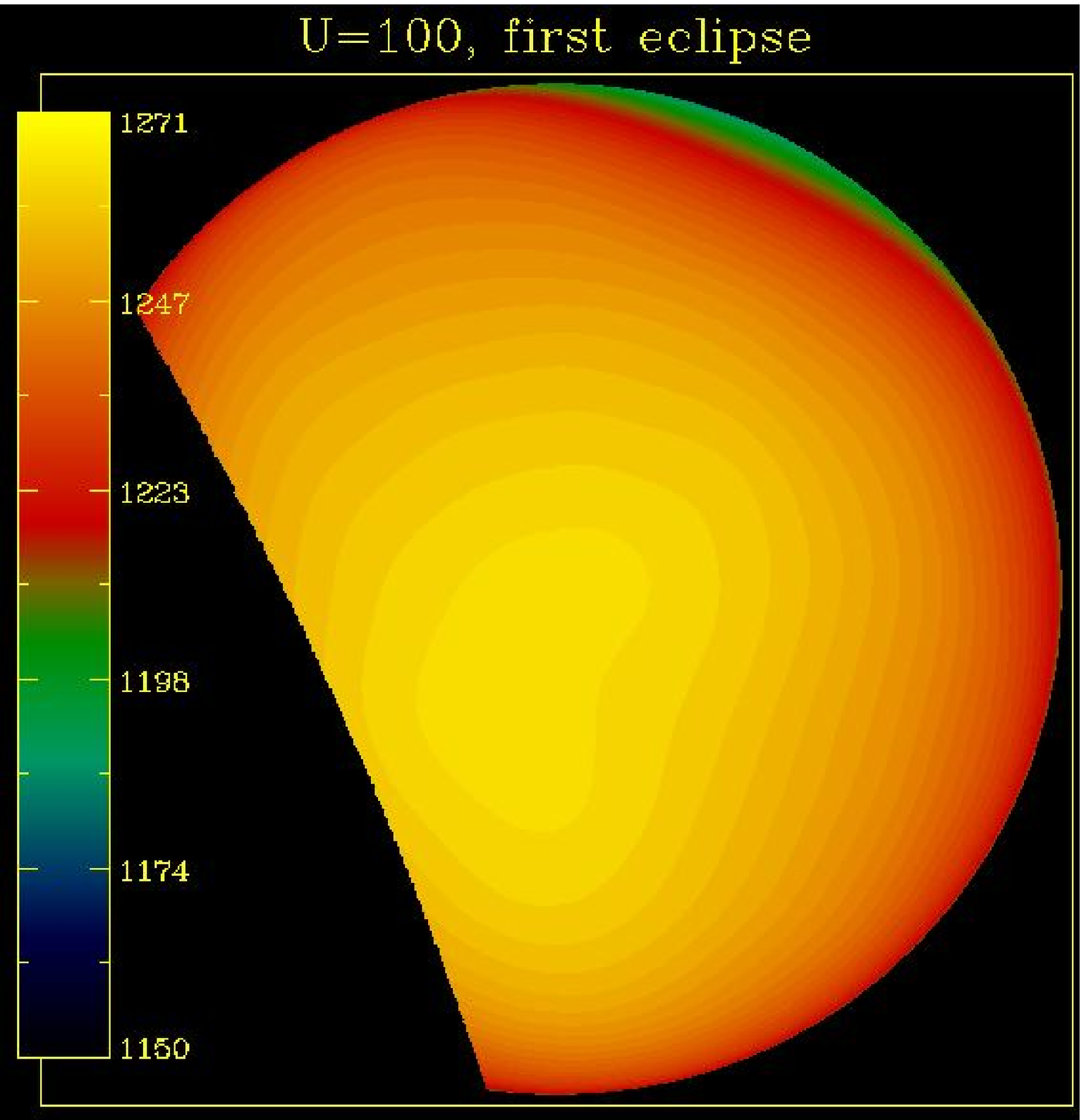}
\includegraphics[width=0.35\textwidth]{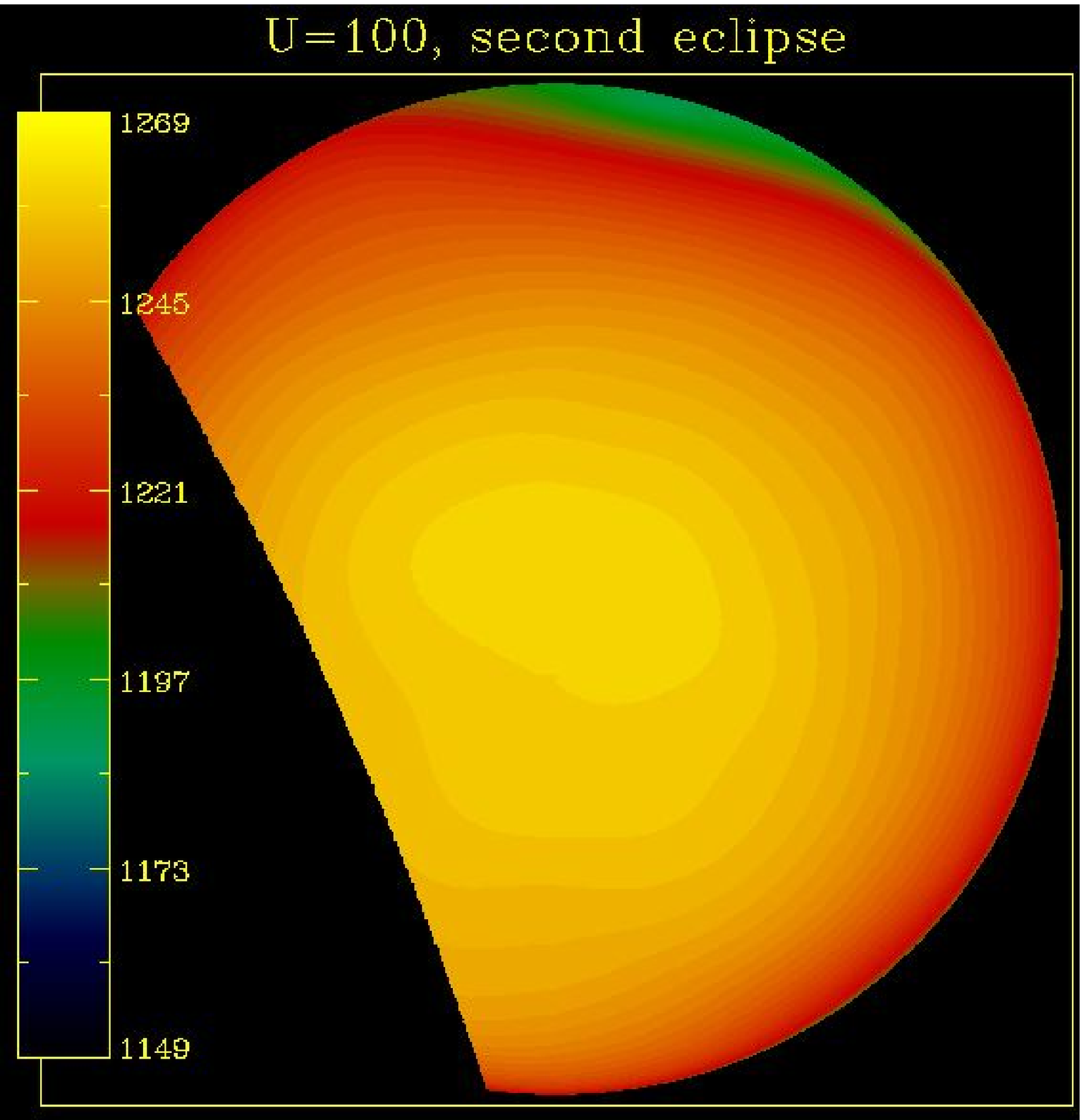}
\caption{Partially eclipsed temperature maps (in K) in a HD209458b
model with a $100$ \ms~ global wind speed, for two successive
eclipses. Little thermal infrared variability is expected.}
\label{fig:models1}
\end{center}
\end{figure}

\begin{figure}
\begin{center}
\includegraphics[width=0.35\textwidth]{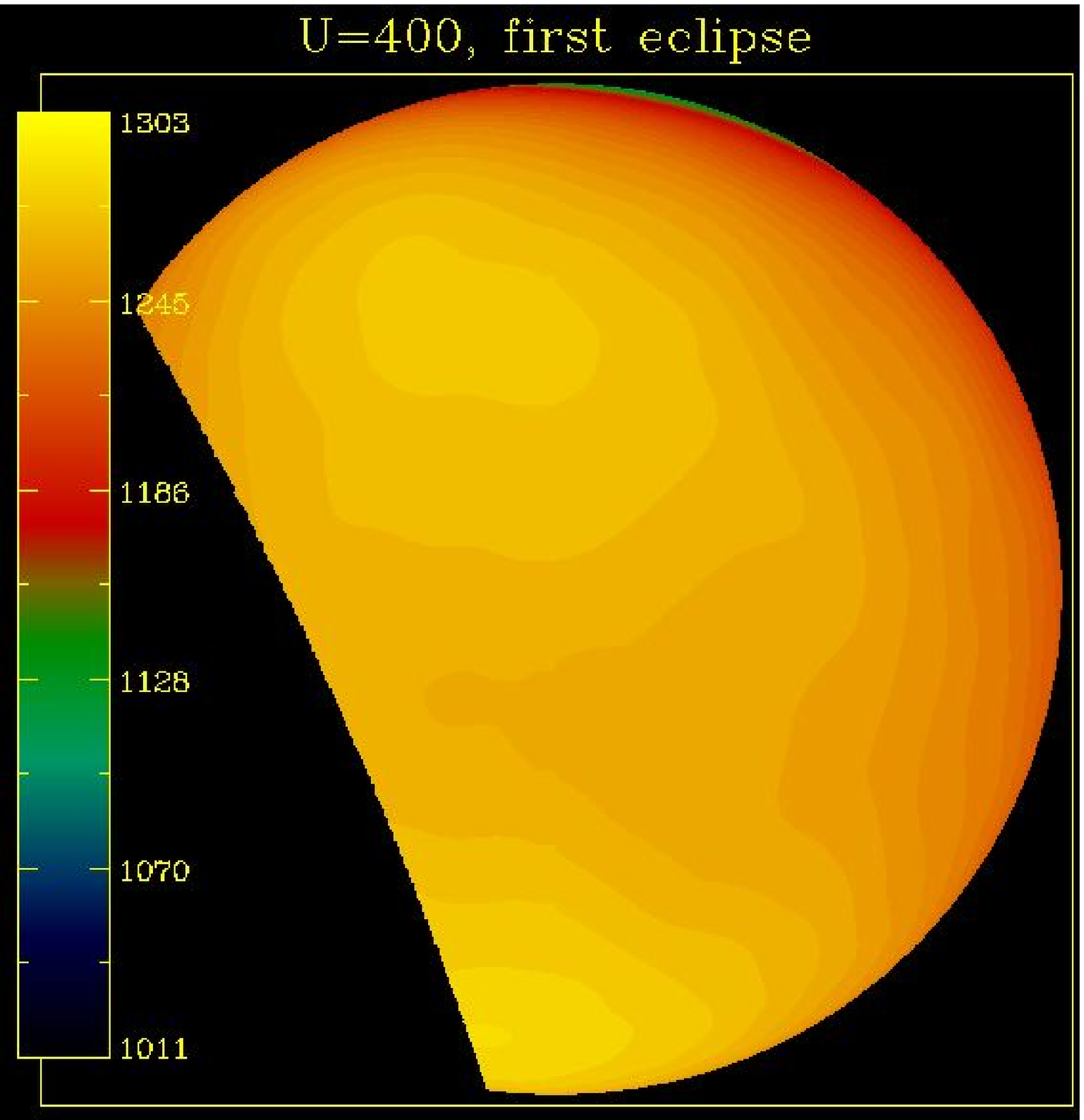}
\includegraphics[width=0.35\textwidth]{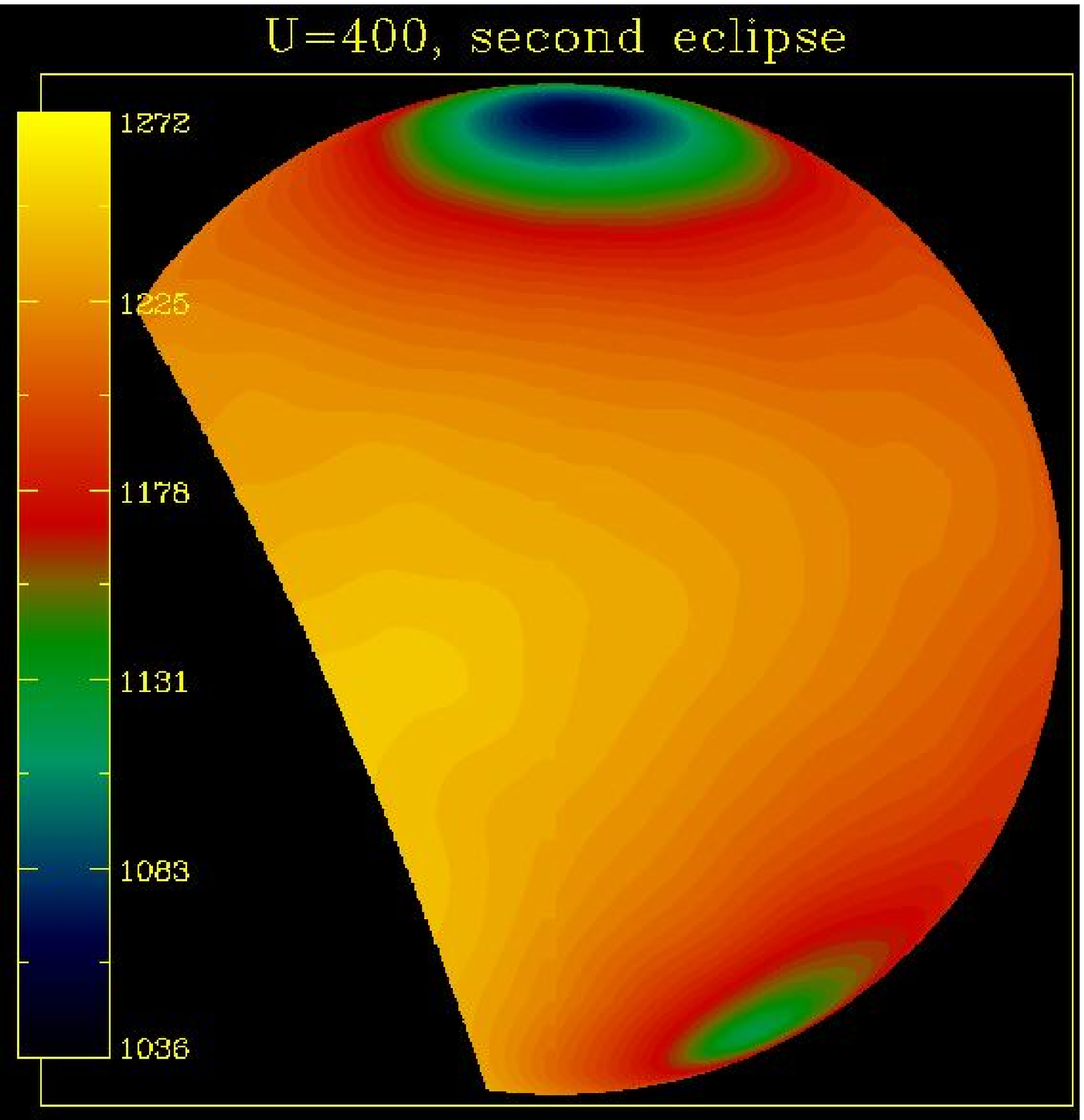}
\caption{Same as Fig.~\ref{fig:models1} in a model with a
$400$~\ms~global wind speed. Significant thermal infrared variability
is expected in this case.} \label{fig:models2}
\end{center}
\end{figure}

Figures~\ref{fig:models1} and~\ref{fig:models2} show snapshots of
orthographically projected, day-side temperature maps in HD209458b
models with $\bar U = 100$ and $400$~\ms, respectively, for two
successive eclipses (i.e. after one HD209458b day). These projections
illustrate how thermal variability in total eclipse depth is expected,
from one eclipse to the next, if large, high contrast temperature
features, associated with moving circulation structures, are present
(in particular, the cyclonic circumpolar vortices most obviously
visible in Fig.~\ref{fig:models2}). Each temperature map in
Figs.~\ref{fig:models1} and~\ref{fig:models2} is shown partially
eclipsed, for the specific geometry of the HD209458 system. According
to these circulation models, cold polar vortices have relatively small
(apparent) areas, so that the magnitude of their contribution to a
variable eclipsed day-side flux is unclear without a detailed
calculation.

\section{Thermal Variability in Eclipse Depth}

Our method to calculate eclipse depths in models like the ones shown
in Figs.~\ref{fig:models1} and~\ref{fig:models2} follows very closely
that used in \citet{Rauscher06} to study partial eclipse diagnostics.
Planet-daily outputs from the circulation simulations provide the
temperature fields used to model successive eclipses.  Accounting for
system specific inclination and geometry, these temperature fields are
orthographically projected onto a 2D disk discretized with ($100$,
$200$) resolution elements in ($r$,$\theta$).  To calculate spectra,
we assume that the vertical temperature profile follows radiative
equilibrium according to the cloudless models of \citet{Seager2005},
under the assumption that the local flow temperature from the
circulation model equals the effective temperature in the radiative
model. We then integrate the spectral emission contributed by each
apparent surface element on the planetary disk, in the global range of
effective temperatures from $700$ to $2000$~K.\footnote{For
temperatures slightly below $700$~K in the $\bar{U}=$800 \ms~models, a
simple linear extrapolation of the spectra is performed.}
Monochromatic fluxes at Earth are then integrated over Spitzer
spectral bands (for IRAC, IRS, and MIPS)\footnote{{\tt
http://ssc.spitzer.caltech.edu/obs/}} to predict the corresponding
successive eclipse depths. {In the future, it will be important to
improve upon this simple treatment by fully incorporating radiative
transfer in circulation models.}

As in \citet{Rauscher06}, we also perform an idealized, bolometric
blackbody analysis, to avoid relying exclusively on model specific
features of the cloudless spectral emission models used. In these
simpler blackbody models, the bolometric flux contributed by each
planetary disk surface element is scaled as the fourth power of the
local photospheric temperature.

Figure~\ref{fig:model_curves}a shows relative variations in successive
eclipse depths predicted by the four HD209458b circulation models,
assuming simple bolometric blackbody emission. Variations are
semi-periodic, as expected from the quasi-periodic motion of the
circumpolar vortices around the planet. As the magnitude of the global
wind speed, $\bar U$, is increased from $100$ to $800$~\ms, leading to
higher contrast motion-induced temperature structures, the amplitude
of thermal infrared variability also increases, from a few \% to
$\sim60$-$70$\%. Figure~\ref{fig:model_curves}b shows eclipse depth
variations for the HD209458b model with $\bar U=400$~\ms~as before,
but this time using the detailed spectral emission models to calculate
contributions in various Spitzer bands.  While the overall variability
scale is similar, it is clear that, by using Spitzer bands, one
preferentially filters emission from a selective range of temperatures
on the planetary disk, which contribute to varying levels of
variability. The effect is substantial and we find that variability is
the strongest in shortest wavelength Spitzer bands, where thermal
contrast from the cold circumpolar vortices is the highest.

We calculated explicit circulation models only for HD209458b but we
can use the general dynamical similarity of hot Jupiter atmospheres
\citep{Menou03} to rescale simply our results for HD189733b and
TrES-1.  Assuming identical dynamics but allowing for different
average atmospheric temperatures, we linearly rescale our model
temperature maps proportionally to the zero albedo,
fully redistributed equilibrium temperature, $T_p$, of the other two
planets \citep[as in][]{Rauscher06}. We find that the eclipse depth
variability properties for HD189733b and TrES-1 do not deviate from
those of HD209458b by more than $\sim10$\%. Increasing planetary
albedos also has little effects on the variability properties, as long
as albedos remain $\ll 1$. Finally, as a matter of generality, we have
checked that arbitrarily varying any system's orbital inclination in
the range 80-90\degrees~ has little effect on its variability
properties.

\section{Detecting Eclipse Depth Variability}

We now address the feasibility of detecting variability in eclipse
depth, at the level predicted by the above models, with Spitzer. Let
us define $\sigma_{ed}$ as the fractional error on the eclipse depth,
that is the ratio of the full (1-$\sigma$) error on the eclipsed flux
to the eclipsed flux itself. Values of $\sigma_{ed}$ for the
existing secondary eclipse measurements \citep[taken
from][]{Charbonneau2005,Deming2005,Deming2006} are listed in bold in
Table~2. These numbers already suggest that eclipse depth variability
at the level of $5$-$20$\% could be detected in these systems.

Our variability models indicate, however, that specific Spitzer bands
may be much more useful than others for eclipse depth variability
detections. We perform simple estimates of $\sigma_{ed}$ errors for
any combination of Spitzer instrument and hot Jupiter system as
follows. We use the same set of system parameters as in Table~1 of
\citet{Rauscher06}.  Assuming that errors on the non-eclipsed flux are
comparatively very small, we write $\sigma_{ed}=\sigma_1 / \sqrt{N}$,
where $\sigma_1$ is the noise per data point in units of the eclipsed
flux and $N$ is the number of single data points collected during a
full eclipse period. $N$ is the ratio of the secondary eclipse
duration to the instrumental cadence (taken from existing
measurements). The eclipse duration time is calculated as $t_{\rm
ec}={2}\sqrt{(R_*+R_p)^2-a^2\sin^2(90^\mathrm{o}-i)}/v$, where
$v=2\pi a/P$ is the planet's orbital velocity, $P$ its orbital period,
$a$ its orbital semi-major axis, $i$ is the orbital inclination, and
$R_*$ and $R_p$ are the stellar and planetary radii, respectively
\citep[see Table~1 of][]{Rauscher06}. We obtain values of $t_{\rm ec}=
1.76$, $3.23$ and $2.55$~hrs for HD189733b, HD209458b and TrES-1,
respectively.

Finally, we extrapolate instrument-specific $\sigma_{ed}$ errors known
for one system to the other two systems of interest by assuming
blackbody emission for both the star and the planet. This results in
the following instrument-specific scaling between systems A and B:
\begin{eqnarray}
\frac{\sigma_{ed}^B}{\sigma_{ed}^A} & = &
\sqrt{\frac{t_{ec}^A}{t_{ec}^B}} \sqrt{\frac{F_{*}^B}{F_{*}^A}}
\left(\frac{F_{p}^A}{F_{p}^B}\right) \nonumber \\ & = &
\sqrt{\frac{t_{ec}^A}{t_{ec}^B}} \left(\frac{d_B}{d_A}\right)
\left(\frac{R_{p}^A}{R_{p}^B}\right)^2
\left(\frac{R_{*}^B}{R_{*}^A}\right) 
\left(\frac{B_\lambda(T_{p}^A)}{B_\lambda(T_{p}^B)}\right)
\sqrt{\frac{B_\lambda(T_{*}^B)}{B_\lambda(T_{*}^A)}}, \nonumber
\end{eqnarray}
\noindent where $d$ is the distance to the system and $B_\lambda$ is
the Planck function evaluated at the central wavelength of the
instrumental band under consideration. $T_*$ is the stellar effective
temperature and $T_p=T_*\sqrt{R_*/2a}$ is the fully redistributed
planetary equilibrium temperature calculated in the small albedo
limit, exactly as in \citet{Rauscher06}. The resulting extrapolated
values of $\sigma_{ed}$ for the three systems of interest are listed
in Table~2.

Repeated eclipse depth measurements with IRAC for HD189733b and
HD209458b appear to be the most likely to succeed in detecting
atmospheric variability, at the few percent level or more. Such
variability measurements would be difficult for
TrES-1. Figure~\ref{fig:model_curves} shows that any pair of
successive eclipses will generally not display the full range of
variability amplitude: more than two eclipse measurements are needed
to sample variability properties adequately. This requirement can be
quantified by calculating distributions of fractional variations in
eclipse depth over series of 2, 3, 4 or more successive eclipse
measurements. Comparing these distributions of eclipse depth
variations to the (1-$\sigma_{ed}$) errors listed in Table~2, we find
that detecting the eclipse depth variability predicted by the
$\bar{U}=$400 or 800 \ms~models at the 2-3 $\sigma$ level requires 3-4
IRAC eclipses (at 4.5 or, slightly better, at 8\micron). A generally
larger number of eclipses is needed to detect the variability
predicted by models with smaller global wind speeds.  Eclipse depth
variations at the level predicted by the $\bar{U}=$100 \ms~model would
be systematically masked by eclipse depth measurement uncertainties,
according to our estimates. Finally, we note that a sufficiently large
number of successive eclipse measurements could reveal the
quasi-periodicity apparent in Fig.~\ref{fig:model_curves}.

\section{Discussion and Conclusion}

We have illustrated, using simple turbulent shallow-layer
circulation models, how eclipse depth variability can be used to
constrain the circulation regime and global wind speeds in hot Jupiter
atmospheres. 

\begin{figure}
\begin{center}
\includegraphics[width=0.45\textwidth]{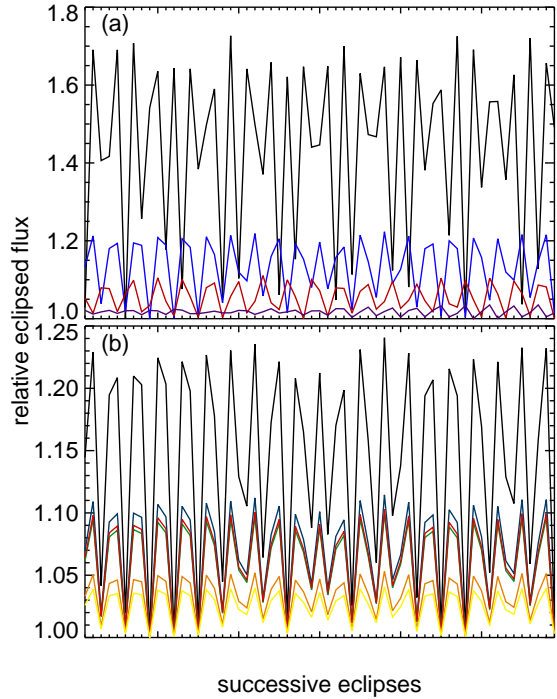}
\caption{Relative flux variations at full eclipse according to
HD209458b circulation models. Discrete points are connected for
clarity and fluxes are normalized to the minimum value in each case.
(a) Variations for simple bolometric blackbody emission, in models
with $\bar{U}$=100 (purple line), 200 (red), 400 (blue), and 800
\ms~(black).  (b) Variations according to detailed spectral emission
models, in various Spitzer bands, for the $\bar{U}$=400 \ms~model of
HD209458b.  From top to bottom, the curves correspond to
3.6\micron~(\emph{black}), 4.5\micron~(\emph{blue}),
8\micron~(\emph{red}), 6\micron~(\emph{green}),
16\micron~(\emph{orange}), and 24\micron~(\emph{yellow}).}
\label{fig:model_curves}
\end{center}
\end{figure}

Clearly, our circulation models and variability predictions are highly
idealized. We have focused on simple thermal diagnostics in models
describing an atmosphere as a single horizontal layer of turbulent
fluid. Issues related to detailed radiative transfer \citep[e.g.,
variation of photospheric height with
wavelength;][]{Iro,Seager2005,HarrUps06}, the presence of
high-altitude haze or the possible existence of clouds could all
seriously affect our conclusions.  {Even strictly within the framework
of our shallow-layer models, we know that the parameterized amplitude
of net radiative forcing, $\eta$, can affect the thermal contrast, and
therefore the detectability, of moving atmospheric structures
\citep[see][for details]{Cho2006}. We have recalculated all our
circulation and variability models with increased values of $\eta=0.1$
and $0.2$.\footnote{In these models with stronger radiative forcing,
an additional $5$-$10$\% contribution to the flow kinetic energy
results from conversion of available potential energy.} We find that
the variability amplitude is reduced by up to a factor of two from the
case with $\eta=0.05$ shown in Fig.~\ref{fig:model_curves}. Values of
$\eta > 0.2$ would further reduce the variability amplitude. In the
future, multi-wavelength phase curve data \citep[e.g.,][]{HarrUps06}
may provide useful observational diagnostics on the relevant value of
$\eta$ for a given planet.}

{Despite these shortcomings}, our results are promising in showing
that eclipse depth variability is a new and potentially powerful tool
to diagnose circulation and wind speeds in hot Jupiter atmospheres. In
the future, as more refined atmospheric models are developed and more
data becomes available, this tool should become increasingly useful in
characterizing hot Jupiter atmospheres.

We thank an anonymous referee for comments that helped improve the
manuscript. This work was supported by NASA contract NNG06GF55G, NASA
Astrobiology Institute contract NNA04CC09A and a Spitzer Theory
grant. q

\clearpage

\begin{deluxetable}{lccc}
\tablewidth{0pt}
\tablecaption{\citet{Cho2006} models under consideration}
\tablehead{
\colhead{}  &  \colhead{} & \multicolumn{2}{c}{Min-Max Temperature (K)} \\
\colhead{}  &  \colhead{$\bar{U}$ (\ms)}  &  \colhead{HD209458b}  & \colhead{HD189733b}
}
\startdata
Model 1  &  100  &  1147-1275 &  956-1062 \\
Model 2  &  200  &  1112-1287 &  926-1072 \\
Model 3  &  400  &  1006-1308 &  838-1088 \\
Model 4  &  800  &  669-1372  &  557-1141 \\
\enddata
\end{deluxetable}

\clearpage

\begin{deluxetable}{lccc}
\tablewidth{0pt}
\tablecaption{Measured (in bold) and estimated values of $\sigma_{ed}$.}
\tablehead{
\colhead{Instrument, wavelength}  &  \colhead{HD 189733b}  &  \colhead{HD 209458b}  &  \colhead{TrES-1}
}
\startdata
IRAC, 4.5 \micron  &  0.018  &  0.039  &  \textbf{0.20} \\
IRAC, 8 \micron  &  0.014  & 0.026  &  \textbf{0.16} \\
IRS, 16 \micron  &  \textbf{0.055} &  0.11  &  0.62  \\
MIPS, 24 \micron  & 0.086  &  \textbf{0.18}  &  0.96  \\
\enddata
\end{deluxetable}

\clearpage

\clearpage

\clearpage

\end{document}